\documentclass[12pt,aps,showpacs,eqsecnum,nofootinbib,floatfix]{revtex4}
\usepackage{bbding}
\usepackage{pifont}
\usepackage{subfigure}
\usepackage{epsfig}
\usepackage{amsmath}
\usepackage{amsfonts}
\usepackage{amssymb}
\usepackage{color}
\usepackage{graphicx}
\usepackage{mathrsfs}
\usepackage{multirow}

\def\be{\begin{equation}}
\def\ee{\end{equation}}
\def\ba{\begin{eqnarray}}
\def\ea{\end{eqnarray}}

\newcommand{\omits}[1]{}

\begin{document}

\title{Phase Transition and Clapeyron Equation of Black Holes in
Higher Dimensional AdS Spacetime}

\author{Hui-Hua Zhao, Li-Chun Zhang, Meng-Sen Ma, Ren Zhao\footnote{corresponding author: Email:zhao2969@sina.com(Ren Zhao)}}

\medskip

\affiliation{Institute of Theoretical Physics, Shanxi Datong University,
Datong 037009, China}
\affiliation{Department of Physics, Shanxi Datong University,
Datong 037009, China}

\begin{abstract}

By Maxwell equal area law we study the phase transition in higher dimensional Anti-de Sitter (AdS)
Reissner-Nordstr\"{o}m (RN) black holes and Kerr black holes in this
paper. The coexisting region of the two phases involved in the phase
transition is found and some coexisting curves are shown in
$P-T$ figures. We also analytically investigate the parameters
which affect the phase transition. To better compare with a general
thermodynamic system, the Clapeyron equation is derived for
the higher dimensional AdS black holes.


\end{abstract}


\maketitle

\bigskip

\section{Introduction}

Black hole is an ideal platform for studying many interesting behaviors in classical
gravity theory. It also can be regarded as a
macroscopic quantum system in that its peculiar thermodynamic
properties and thermodynamic quantities, as entropy, temperature, and
its gravity holographic nature are quantum mechanical intrinsically.
Hence research of black hole thermodynamics is an important
channel for studying quantum gravity \cite{JDB1,JDB2,Hawking1,Lu,ZhaoZhang}.
Although thermodynamics of black hole has been researched for many
years, the exact statistical interpretation for thermodynamics state
of black hole is not clear. So the thermodynamics of black hole
remains an important subject.

It has been found that a black hole possesses not only standard
thermodynamic quantities but abundant phase structures and critical
phenomena, such as those involved in Hawking-Page phase transition\cite{Hawking2},
similar to the ones of a general thermodynamic system. Even more
interesting is that the studies on the charged black
holes show they may have an analogous phase transition with that of
van der Waals-Maxwell liquid-gas\cite{Cham1,Cham2}. Recently
critical behaviors and phase transition of black holes have been extensively investigated by considering
the cosmological constant as thermodynamic
pressure\cite{Kubi1,Dolan,Guna,Cvet,Kastor:2009wy},$P=-\frac{1}{8\pi}\Lambda
=\frac{(d-1)(d-2)}{16\pi l^2}$. People have been trying to construct a complete liquid-gas analogue
system for black holes.

In ref.\cite{Kubi1,Guna,Banerjee1,Banerjee2,Banerjee3,Banerjee4,Banerjee5,Banerjee6},
the phase transition and critical behaviors of some AdS black holes have been studied,
which exhibits several different phase transition behaviors. On the basis of the thermodynamic volume of dS
spacetime given in ref.\cite{Dolan,Kastor:2009wy,Alta1}, considering the
connection between black hole horizon and cosmological horizon,
we derived the effective temperature and
effective pressure in some dS spacetimes in our previous works\cite{Zhaohh,Zhao1,Mams1,Zhanglc}. The relation among the
effective thermodynamic quantities was investigated, which shows
the critical phenomena like that in van der Waals
liquid-gas system. Using Ehrenfest scheme, the second order phase
transition was proved to exist in AdS spacetime black hole at the
critical states\cite{Banerjee1,Banerjee4,Mo1,Mo2,Mo3,Mo4,Mo5,Mo6,Mams2,Lala}. Alike conclusion
has been reached by investigating thermodynamics and state space
geometry of black holes in ref.\cite{Banerjee1,Lala,Wei1,Suresh,Mans,Thar,Niu}.
The phase transition behaviors of black holes is found to be related to not only the spacetime metric
but also the theory of gravity or other factors\cite{Mams2,Mams3,Mams4,Gim,Posh,Arci}.

Although some encouraging results have been achieved about the
thermodynamics of black holes in AdS and dS spacetimes, many
thermodynamic properties need to be investigated more specifically.
It is significant to study the critical behaviors and the process of
noncritical phase transition in detail. The analyses on the $P\sim
v$ relation of some black holes in AdS spacetimes show a negative
pressure appear as the temperature is below a certain value and a
thermodynamic unstable region exists with $\partial P/\partial
v>0$\cite{Kubi1,Guna,Zhao2,Zhao3,Alta2,Cai,Zou1,Zou2,Fras,Li,Wei2,Kubi2,Xu,Liu,Hendi}.
Negative pressure, that is, positive $\Lambda$, doesn't correspond
to AdS spacetime. The situation is found in var der Waals-Maxwell
liquid-gas system, where it is solved by the remarkable Maxwell's
equal area law\cite{Spal1,Zhaojx,Spal2,ZhangAHEP}.

In this paper Maxwell's equal area law is extended to analyze the phase
transition in the RN-AdS black holes and
the Kerr-AdS black holes. It can help to resolve the puzzles about the
negative pressure and the unstable region. By Maxwell's equal area law we
discover the possible two phase coexistence curves in the process of phase transition
and the boundary curves of the coexistence region in $P-v$ plots. We expect to
provide some particular information about phase transition and properties of black
hole thermodynamic system in AdS spacetimes to contribute to search for more
stable black holes and to explore properties of quantum gravity.

The paper is arranged as follow: the thermodynamic quantities of
the $d$-dimensional RN AdS black hole and the $d$-dimensional Kerr AdS black hole
are introduced firstly in section 2 and in section 3 respectively. Then the
phase transition of the black holes are studied by Maxwell equal area law,
and the effect of the dimension and the thermodynamic quantities, as electric
charge in section 2 and angular momentum in section 3, on the phase
transition is analyzed . In section 4, we make some discussion. (we use the units $G_d =\hbar =k_B =c=1)$

\section{The charged black holes}

\subsection{Thermodynamics}

Reissner-Nordstrom black holes are characterized by the spherically
symmetry and electrical nature. The solution for $d$-dimensional
RN-AdS spacetime with a negative cosmological constant, $\Lambda
=-(d-1)(d-2)/2l^2$, is defined by the line
element\cite{Cham1}
\be
\label{eq1}
ds^2=-fdt^2+f^{-1}dr^2+r^2d\Omega_{d-2}^2 ,
\ee
where
\be
\label{eq2}
f=1-\frac{m}{r^{d-3}}+\frac{q^2}{r^{2(d-3)}}+\frac{r^2}{l^2}.
\ee
The ADM mass and the electric charge have been defined as ($G_d =1)$,
\be
\label{eq3}
M=\frac{\omega _{d-2} (d-2)}{16\pi }m,
\quad
Q=\frac{\omega _{d-2} \sqrt {2(d-2)(d-3)} }{8\pi }q,
\end{equation}
in which the volume of the unit $d$-sphere $\omega _d $ can be expressed as
\be
\label{eq4}
\omega _d =\frac{2\pi ^{(d+1)/2}}{\Gamma \left( {\textstyle{{d+1} \over 2}}
\right)}.
\ee
The corresponding Hawking temperature, entropy, and electric potential of
the system are defined as
\be
\label{eq5}
T=\frac{d-3}{4\pi r_+ }\left( {1-\frac{q^2}{r_+^{2(d-3)}
}+\frac{d-1}{d-3}\frac{r_+^2 }{l^2}} \right),
\quad
S=\frac{\omega _{d-2} r_+^{d-2} }{4},
\quad
\Phi =\sqrt {\frac{d-2}{2(d-3)}} \frac{q}{r_+^{d-3} }.
\ee
In our consideration we interpret the cosmological constant $\Lambda$
as a thermodynamic pressure $P$\cite{Kastor:2009wy},
\be
\label{eq6}
P=-\frac{\Lambda }{8\pi }=\frac{(d-1)(d-2)}{16\pi l^2}.
\ee
The corresponding conjugate quantity, the thermodynamic volume, is
given by\cite{Kastor:2009wy}
\be
\label{eq7}
V=\frac{\omega _{d-2} r_+^{d-1} }{d-1}
\ee
with $r_+ $ being the position of black hole horizon determined from $f(r_+
)=0$. Combining Eqs. (\ref{eq5}) and (\ref{eq6}), we can obtain
\be
\label{eq8}
P=\frac{T(d-2)}{4r_+ }-\frac{(d-3)(d-2)}{16\pi r_+^2
}+\frac{q^2(d-3)(d-2)}{16\pi r_+^{2(d-2)} }.
\ee
Comparing (\ref{eq8}) with the Van der Waals equation, we conclude that the
specific volume $v$ should be identified with the horizon radius of the
black hole as\cite{Kubi1,Guna}
\be
\label{eq9}
v=\frac{4r_+ l_p^{d-2} }{d-2}.
\ee
In geometric units we have
\be
\label{eq10}
r_+ =kv,
\quad
k=\frac{d-2}{4},
\ee
and the equation of state reads
\be
\label{eq11}
P=\frac{T}{v}-\frac{(d-3)}{\pi (d-2)v^2}+\frac{q^2(d-3)}{4\pi
v^{2(d-2)}k^{2d-5}}.
\ee
In Fig.1, the isotherms are plotted in $P-v$ diagrams for different $d$ and
$q$. It shows there are thermodynamic unstable regions with $\partial
P/\partial v>0$ in the extended phase space, and the situation of negative pressure
appears when $T<\widetilde{T}$. The isotherm corresponding to
$T=\widetilde{T}$ is tangent to horizontal axis with $P=0$, in which $\tilde
{T}$ can be obtained by

\begin{figure}[!htbp]
\center{\subfigure[~$d=4$; $q=2$,\;$3$,\;$4$] {
\includegraphics[angle=0,width=5cm,keepaspectratio]{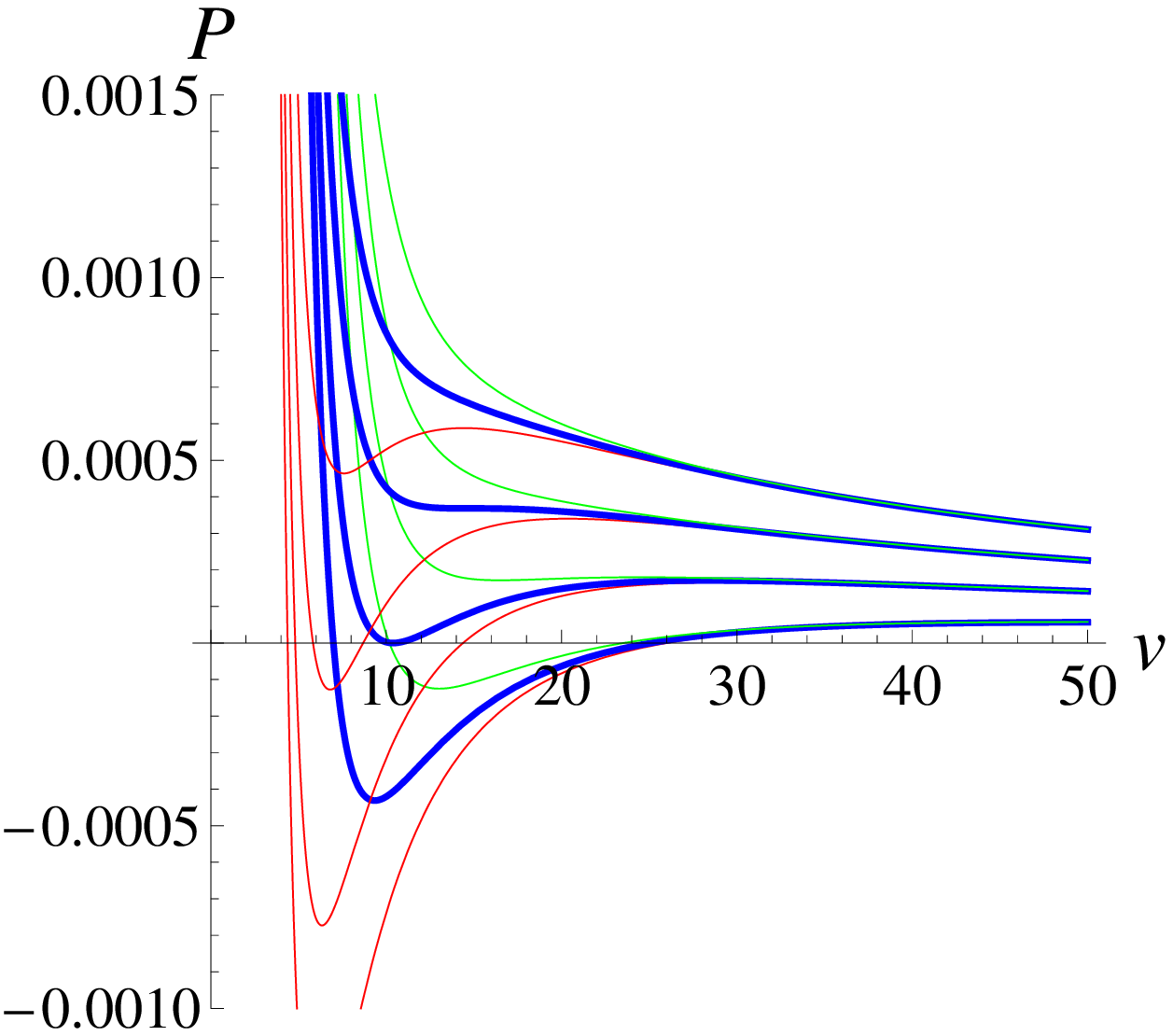}}
\subfigure[~$d=8$; $q=1$,\;$3$,\;$5$] {
\includegraphics[angle=0,width=5cm,keepaspectratio]{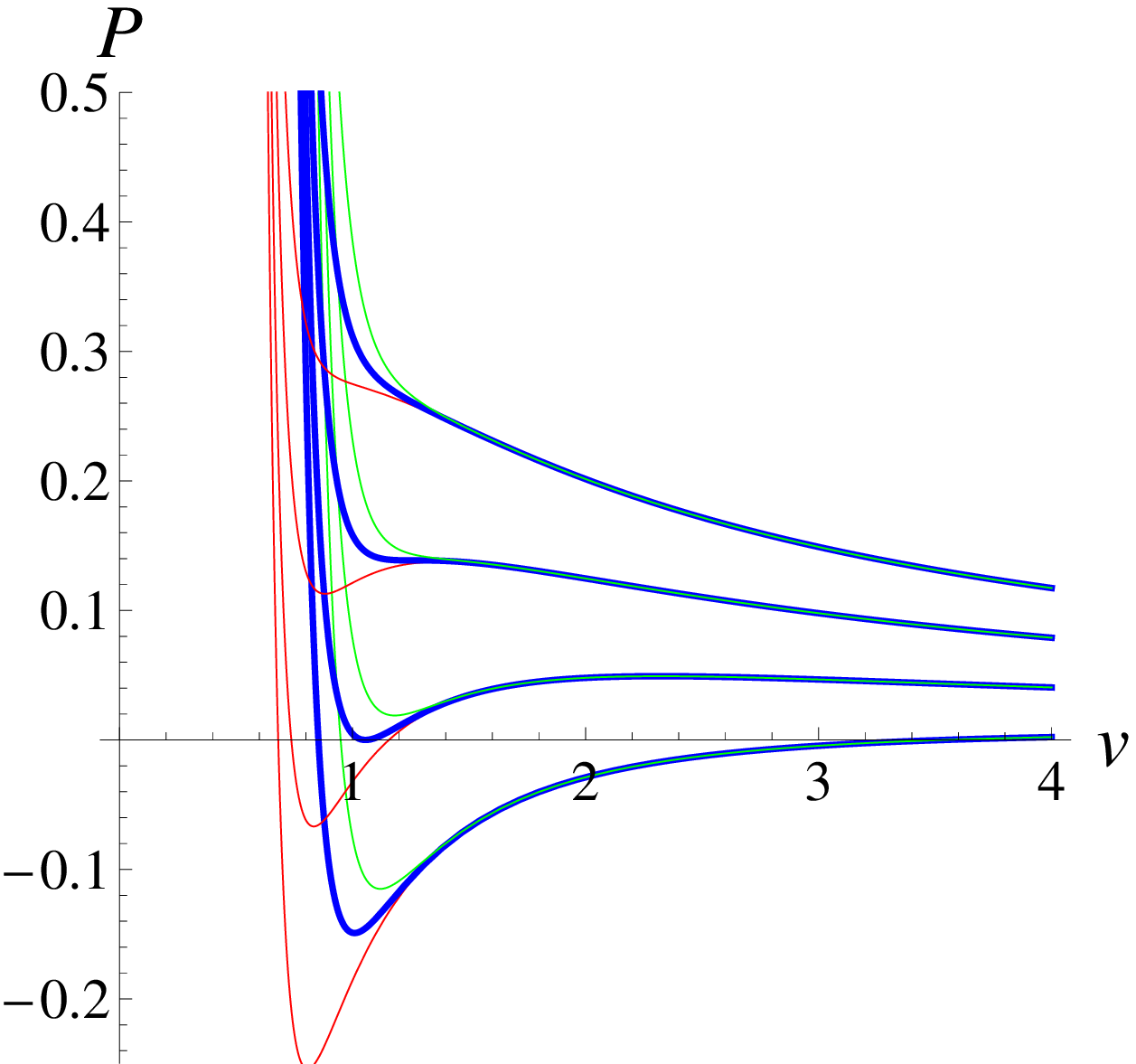}}
\subfigure[~$d=10$; $q=1$,\;$3$,\;$5$] {
\includegraphics[angle=0,width=5cm,keepaspectratio]{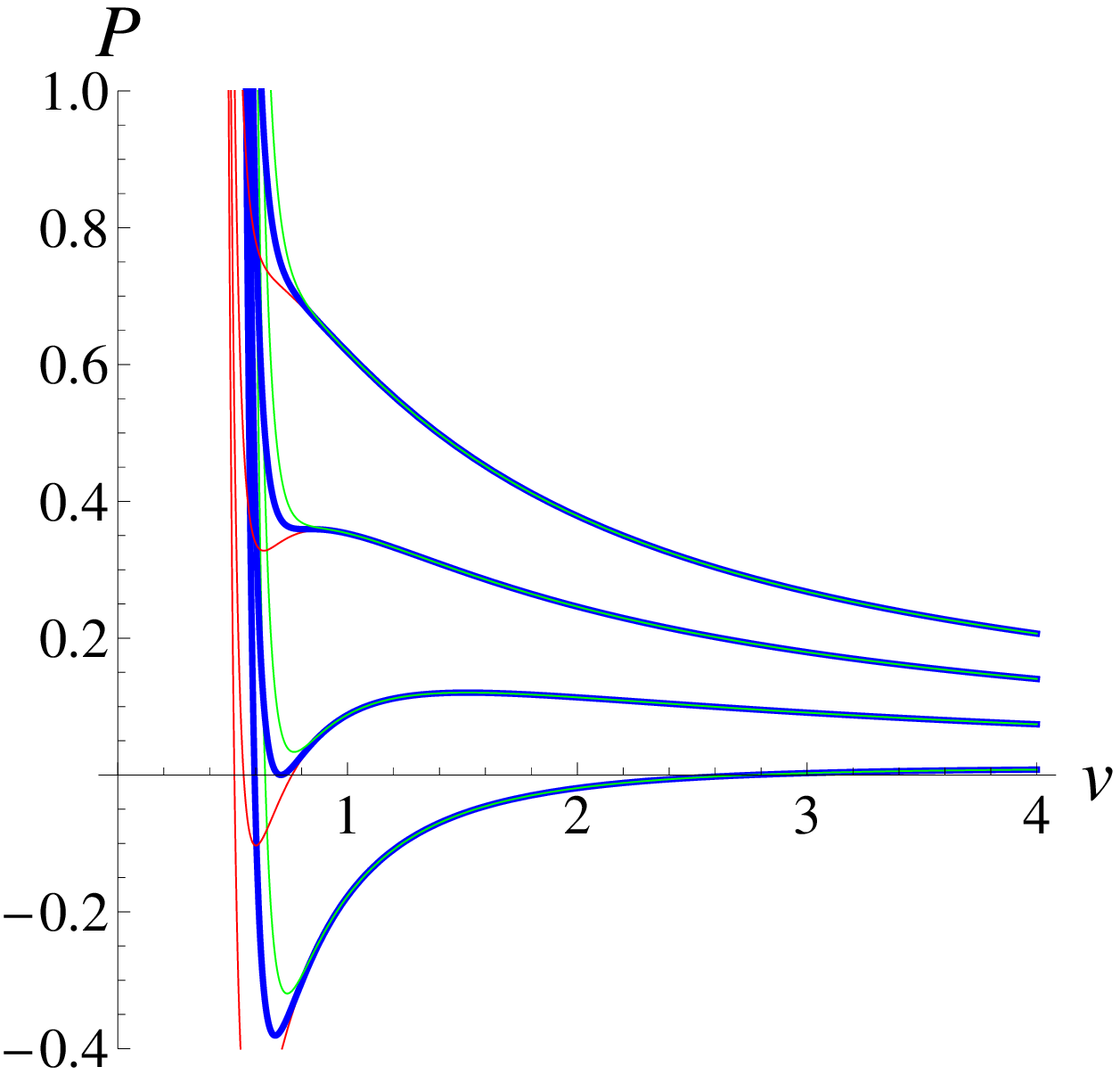}}
\caption[]{\it Isotherms in $P-v$ diagrams. In each diagram the
three groups of curves correspond to the three given values of $q$,
the lower ones (red curves) correspond to the smallest $q$, the
middle ones (blue curves) meet with the medium $q$, and the upper
ones (green curves) are with the biggest $q$. The five curves in
every group represent five different temperatures, the lower curve
the lower temperature.}} \label{Pv1}
\end{figure}

\be
\label{eq12}
\frac{\widetilde{T}(d-2)}{4r_+ }-\frac{(d-3)(d-2)}{16\pi r_+^2
}+\frac{q^2(d-3)(d-2)}{16\pi r_+^{2(d-2)} }=0,
\quad
\left( {\frac{\partial P}{\partial v}} \right)_{\tilde {T}} =0.
\ee
\be
\label{eq13}
\widetilde{T}=\frac{2(d-3)^2}{\pi (d-2)(2d-5)v_0 }.
\ee

\subsection{Maxwell equal area law and two phase equilibria}

The state equation of the $d$-dimensional RN-AdS black hole is exhibited in
the $P-v$ diagrams in Fig.1, which show the existence of negative
pressure and thermodynamic unstable region with $\partial P/\partial v>0$,
which makes a thermodynamic system contract and expand automatically in classical thermodynamics.
The case occurs in van der Waals equation, where the problems have been resolved by
reference to some practical process of phase transition in real fluid and by
Maxwell equal area law.

Using Maxwell equal area law to deal with the state equation of the
$d$-dimensional RN-AdS black hole, the two-phase equilibrium lines, where
the two phases coexist, are acquired.

Take $T_c $ as critical temperature of the $d$-dimensional RN-AdS black hole,
and suppose at temperature $T_0 $ ($T_0 \le T_c )$ the boundary state parameters for
two-phase coexistence are $(v_1 ,P_0 )$ and $(v_2 ,P_0 )$. According to
Maxwell equal area law,
\be
\label{eq14}
P_0 (v_2 -v_1 )=\int\limits_{v_1 }^{v_2 } {Pdv} ,
\ee
we can get
\be
\label{eq15}
P_0 (v_2 -v_1 )
=T_0 \ln \left( {\frac{v_2 }{v_1 }} \right)+\frac{(d-3)}{\pi (d-2)}\left(
{\frac{1}{v_2 }-\frac{1}{v_1 }} \right)-\frac{q^2(d-3)}{4(2d-5)\pi
k^{2d-5}}\left( {\frac{1}{v_2^{2d-5} }-\frac{1}{v_1^{2d-5} }} \right),
\ee
\be
\label{eq16}
P_0 =\frac{T_0 }{v_1 }-\frac{(d-3)}{\pi (d-2)v_1^2 }+\frac{q^2(d-3)}{4\pi
v_1^{2(d-2)} k^{2d-5}},
\quad
P_0 =\frac{T_0 }{v_2 }-\frac{(d-3)}{\pi (d-2)v_2^2 }+\frac{q^2(d-3)}{4\pi
v_2^{2(d-2)} k^{2d-5}}.
\ee
Set $x:=v_1 /v_2 $, from Eqs. (\ref{eq15}) and (\ref{eq16}), we can obtain
\be
\label{eq17}
v_2^{2d-6} =q^2\frac{y_1 (x)}{y_2 (x)},
\ee
where
\[
y_1 (x)=\frac{(d-3)}{2k^{2d-5}}\left( {(1-x^{2d-4})\ln x+\left[
{\frac{(1-x^{2d-5})}{2d-5}+1-x^{2d-5}} \right](1-x)} \right),
\]
\be
\label{eq18}
y_2 (x)=\frac{2(d-3)}{(d-2)}x^{2d-6}(1-x)\left( {(1+x)\ln x+2(1-x)}
\right).
\ee
While $x\to 1$, from (\ref{eq17}) we get the critical specific volume
\be
\label{eq19}
v_c =\frac{1}{k}\left[ {q^2(d-2)(2d-5)} \right]^{1/(2d-6)}.
\ee
According to (\ref{eq16}),
\be
\label{eq20}
T_0 v_2^{2d-5} x^{2d-5}=\frac{(d-3)}{\pi (d-2)}v_2^{2d-6}
x^{2d-6}(1+x)-\frac{q^2(d-3)}{4\pi k^{2d-5}}\frac{(1-x^{2(d-2)})}{1-x},
\ee
and the critical temperature and critical pressure are obtained as $x\to 1$,
\be
\label{eq21}
T_c =\frac{(d-3)^2}{\pi v_c k(2d-5)},
\ee
\be
\label{eq22}
P_c =\frac{(d-3)^2}{16\pi v_c^2 k^2}.
\ee
These critical thermodynamic quantities are consistent with those
in Ref.\cite{Guna}.

Combining (\ref{eq17}), (\ref{eq20}) and (\ref{eq21}), as $T_0 =\chi T_c $ with $\chi \le 1$,
we can get
\[
\chi \frac{(d-3)x^{2d-5}}{(2d-5)[(d-2)(2d-5)]^{1/[2(d-3)]}}\left( {\frac{y_1
(x)}{y_2 (x)}} \right)^{(2d-5)/(2d-6)}
\]
\be
\label{eq23}
=\frac{(1+x)}{(d-2)}x^{2d-6}\left( {\frac{y_1 (x)}{y_2 (x)}}
\right)-\frac{1}{4k^{2d-5}}\frac{(1-x^{2(d-2)})}{1-x}.
\ee
For a fixed $\chi $, i.e. a fixed $T_0 $, we can evaluate $x$ with (\ref{eq23}),
and then according to (\ref{eq17}) and (\ref{eq16}), the $v_2 $, $v_1 $ and $P_0 $ are
specified while$q$ is determined. Join the points $(v_1 ,P_0 )$ and $(v_2 ,P_0 )$ on the
isotherm with $T=T_0$ in the $P-v$ diagram, which generate an isobar
representing the process of phase transition like that of van der Waals
system. Fig.2 shows the isobars with solid (red) straight lines and the boundary of the region of
two phase coexistence by the dot-dashed (green) curve. In Fig.2 each
combined solid line simulates the process of thermodynamic state change and
phase transition of $d$-dimensional RN-AdS black hole at a certain
temperature. The phase transition process becomes shorter as temperature
goes up until it turns into a single point at a certain temperature, which
is the critical temperature, and the point corresponds to critical state of
$d$-dimensional RN-AdS black hole.

\begin{figure}
  \centering
  \includegraphics[width=3in]{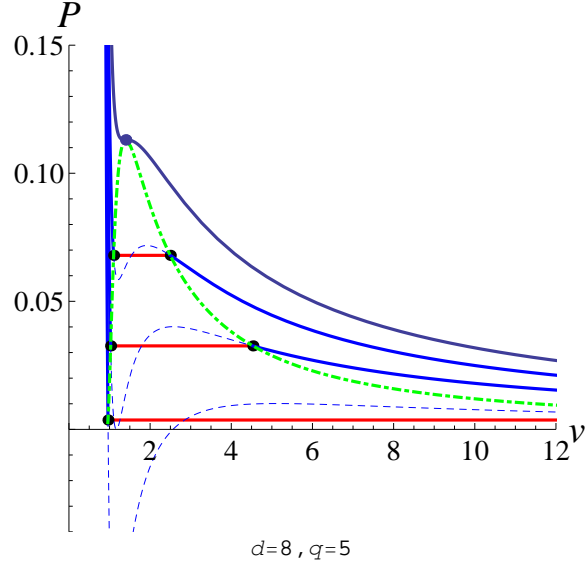}\\
  \caption{\it The simulated phase transition(red solid lines) and the boundary of
two phase coexistence (green dot-dashed curve) on the base of isotherms in
$P-v$ diagram for RN-AdS black hole with $d=8$, $q=5$.}\label{coexist1}
\end{figure}

When taking $\chi =0.1,\;0.3,\;0.5,\;0.7,\;0.9$, we have calculated the
quantities $x$, $v_2 $, $P_0 $ as $q$=0.5, 1, 1.5 in the spacetimes with the dimension
$d=4,\;8,\;10$ respectively. The results are shown in Table 1.

\begin{table}[htbp]
\caption{\it The parameters of two phase coexistence for RN-AdS black
hole.}
\begin{center}
\begin{tabular}{|p{20pt}|p{20pt}|p{40pt}|p{40pt}|p{40pt}|p{40pt}|p{40pt}|p{40pt}|p{40pt}|p{40pt}|p{40pt}|}
\hline & & \multicolumn{3}{|p{120pt}|}{$d=4$} &
\multicolumn{3}{|p{120pt}|}{$d=8$} &
\multicolumn{3}{|p{120pt}|}{$d=10$}  \\

\hline $q$&  $\chi $& $x$& $v_2 $& $P_0 $& $x$& $v_2 $& $P_0 $& $x$&
$v_2 $&
$P_0 $ \\

\hline \raisebox{-6.00ex}[0cm][0cm]{$0.5$}& 0.1& 1.79E{-6}&
5.74E{5}& 1.51E{-8}& 1.86E{-4}& 3.17E{3}& 1.72E{-5}& 3.26{-4}&
1.40E{3}&
5.80E{-5} \\

\cline{2-11}
 & 0.3& 0.00712& 155.& 1.61E{-4}& 0.0378& 16.0& 0.00919& 0.0470& 9.99& 0.0217 \\

\cline{2-11}
 & 0.5& 0.0496& 24.3& 0.00151& 0.149& 4.24& 0.0497& 0.173& 2.80& 0.110 \\

\cline{2-11}
 & 0.7& 0.149& 9.15& 0.00475& 0.324& 2.06& 0.123& 0.361& 1.40& 0.265 \\

\cline{2-11}
 & 0.9& 0.387& 4.36& 0.00995& 0.595& 1.23& 0.225& 0.635& 0.862& 0.477 \\

\hline \raisebox{-6.00ex}[0cm][0cm]{$1$}& 0.1&1.79E{-6}& 1.15E{6}&
3.77E{-9}& 1.86E{-4}& 3.64E{3}& 1.31E{-5}& 3.26E{-4}& 1.55E{3}&
4.76E{-5} \\

\cline{2-11}
 & 0.3& 0.00712& 310.& 4.03E{-5}& 0.0378& 18.4& 0.00696& 0.0470& 11.0& 0.0178 \\

\cline{2-11}
 & 0.5& 0.0496& 48.7& 3.78E{-4}& 0.149& 4.87& 0.0377& 0.173& 3.09& 0.0903 \\

\cline{2-11}
 & 0.7& 0.149& 18.3& 0.00119& 0.324& 2.37& 0.0933& 0.361& 1.55& 0.218 \\

\cline{2-11}
 & 0.9& 0.387& 8.71& 0.00249& 0.595& 1.42& 0.170& 0.635& 0.951& 0.391 \\

\hline \raisebox{-6.00ex}[0cm][0cm]{$1.5$}& 0.1& 1.79E{-6}&
1.72E{6}& 1.68E{-9}& 1.86E{-4}& 3.95E{3}& 1.11E{-5}& 3.26E{-4}&
1.64E{3}&
4.24E{-5} \\

\cline{2-11}
 & 0.3& 0.00712& 465.& 1.79E{-5}& 0.0378& 20.0& 0.00592& 0.0470& 11.7& 0.0159 \\

\cline{2-11}
 & 0.5& 0.0496& 73.0& 1.68E{-4}& 0.149& 5.28& 0.0320& 0.173& 3.28& 0.0804 \\

\cline{2-11}
 & 0.7& 0.149& 27.5& 5.27E{-4}& 0.324& 2.57& 0.0793& 0.361& 1.642& 0.194 \\

\cline{2-11}
 & 0.9& 0.387& 13.1& 0.00111& 0.595& 1.54& 0.145& 0.635& 1.01& 0.348 \\
\hline
\end{tabular}
\label{tab1}
\end{center}
\end{table}

From Table 1, it can be seen that $x$ is unrelated to $q$, but
incremental with the increase of $\chi$/$d$  at certain $d$/$\chi$.
$v_2$ decreases with the increasing $\chi$/$d$ and increases with the
incremental $q$ when the other parameters are fixed respectively . $P_0 $ increases
with the incremental $\chi$/$d$ and decreases with the increasing $q$ with others determined parameters
respectively.

\subsection{$P-T$ curves of two phase coexistence}

From isothermal $P-v$ curves in Fig.1, we know that when temperature
$T<\tilde {T}$, the negative pressure section on the isotherm emerges and
becomes larger with lower temperature. From Fig.2, we can see that the lower temperature, the smaller
the value of $P_0 $, which also can be seen in Table 1. The doubt is whether $P_0$
is negative when temperature is low enough.

Considering (\ref{eq17}), (\ref{eq16}) can be rewritten as
\[
P_0 q^{2/(d-3)}\left( {\frac{y_1 (x,)}{y_2 (x)}} \right)^{(d-2)/(d-3)}=\chi
\frac{(d-3)^2}{\pi (2d-5)[(d-2)(2d-5)]^{1/[2(d-3)]}}\left( {\frac{y_1
(x)}{y_2 (x)}} \right)^{(2d-5)/(2d-6)}
\]
\be
\label{eq24}
-\frac{(d-3)}{\pi (d-2)}\left( {\frac{y_1 (x,)}{y_2 (x)}}
\right)+\frac{(d-3)}{4\pi k^{2d-5}}
\ee
From (\ref{eq23}), (\ref{eq24}) and $T_0 =\chi T_c $, one can get the relation between
$P_0 $ and $T_0 $, which is shown in Fig.3. Fig.3 exhibits the $P-T$ phase
diagrams at fixed $q$ and $d$, in which the curves represent the states of two phase
coexistence and the terminal points are the critical points. From
Fig.3 it can be seen that the influence of the electric charge $q$ and spacetime
dimension $d$ on the phase diagrams, however pressure $P_0 $ tends toward
zero with decreasing temperature $T_0 $ for all of the fixed $q$
and $d$ cases. That the pressure $P_0 $ is always positive means Maxwell equal area law
is appropriate to resolve the doubts about negative pressure and unstable
states in phase transition of the $d$-dimensional RN-AdS black hole. In Fig.4,
the $P_0 q^{2/(d-3)}-\chi $ connection is depicted at different spacetime dimension $d$, which
explicitly exhibits the effect of the dimension $d$ on the distribution of the
black hole phases.

\begin{figure}[!htbp]
\center{\subfigure[~$d=4$; $q=1$,\;$3$,\;$5$] {
\includegraphics[angle=0,width=5cm,keepaspectratio]{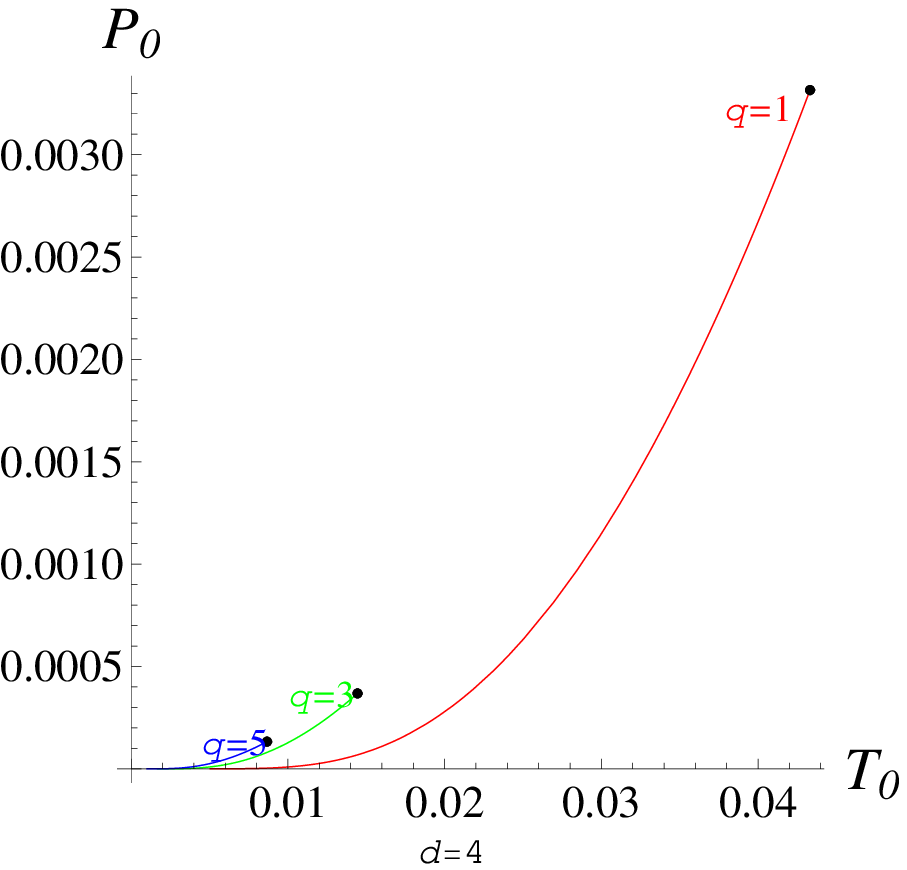}}
\subfigure[~$d=8$; $q=1$,\;$3$,\;$5$] {
\includegraphics[angle=0,width=5cm,keepaspectratio]{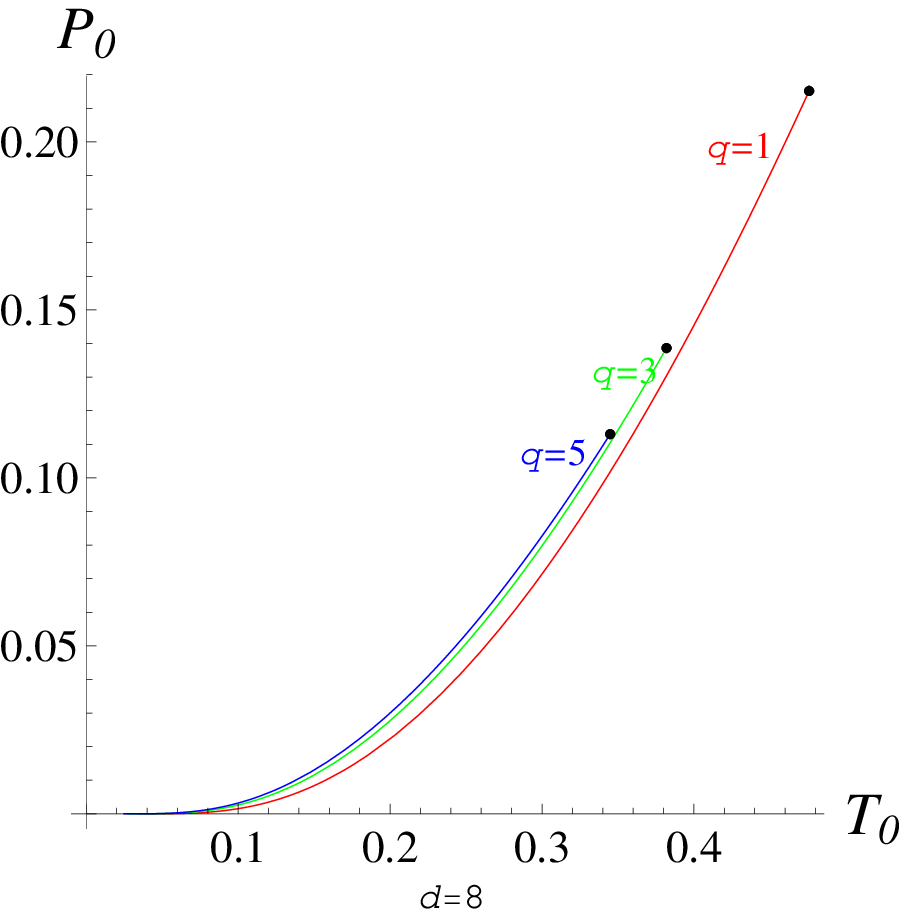}}
\subfigure[~$d=10$; $q=1$,\;$3$,\;$5$] {
\includegraphics[angle=0,width=5cm,keepaspectratio]{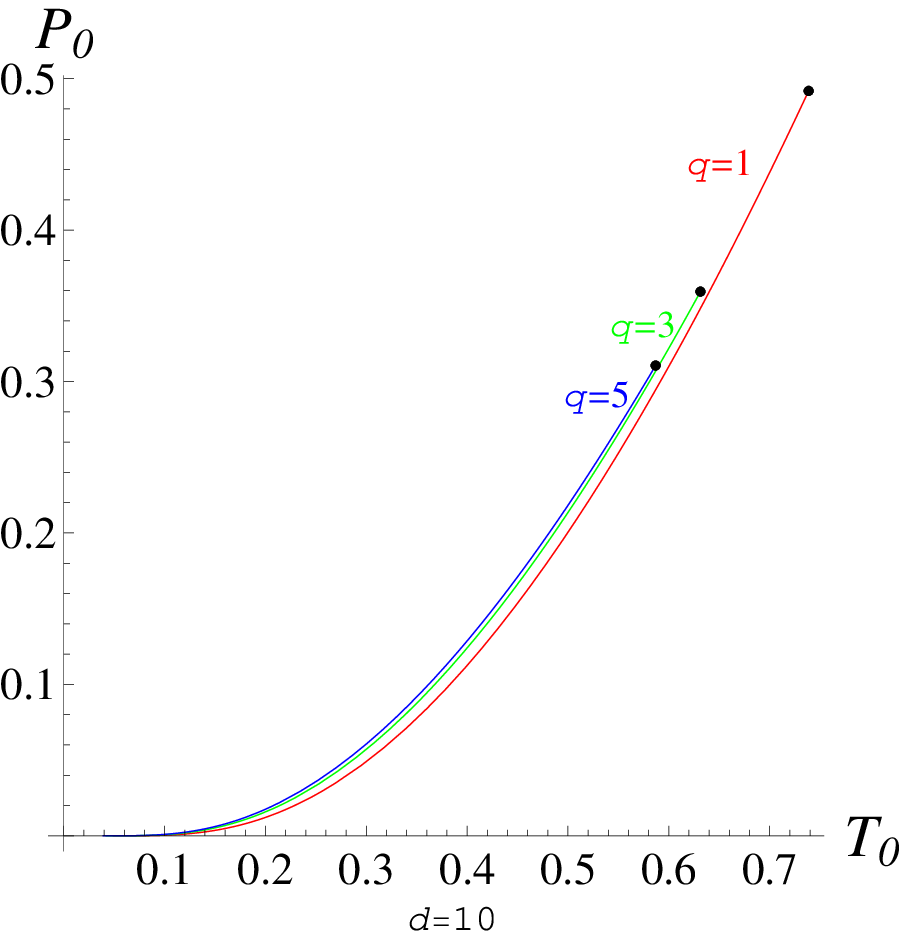}}
\caption[]{\it $P-T$ phase diagrams at fixed $q$ and $d$ for
$d$-dimensional RN-AdS black hole.}} \label{PT1}
\end{figure}

\begin{figure}
  \centering
  \includegraphics[width=3in]{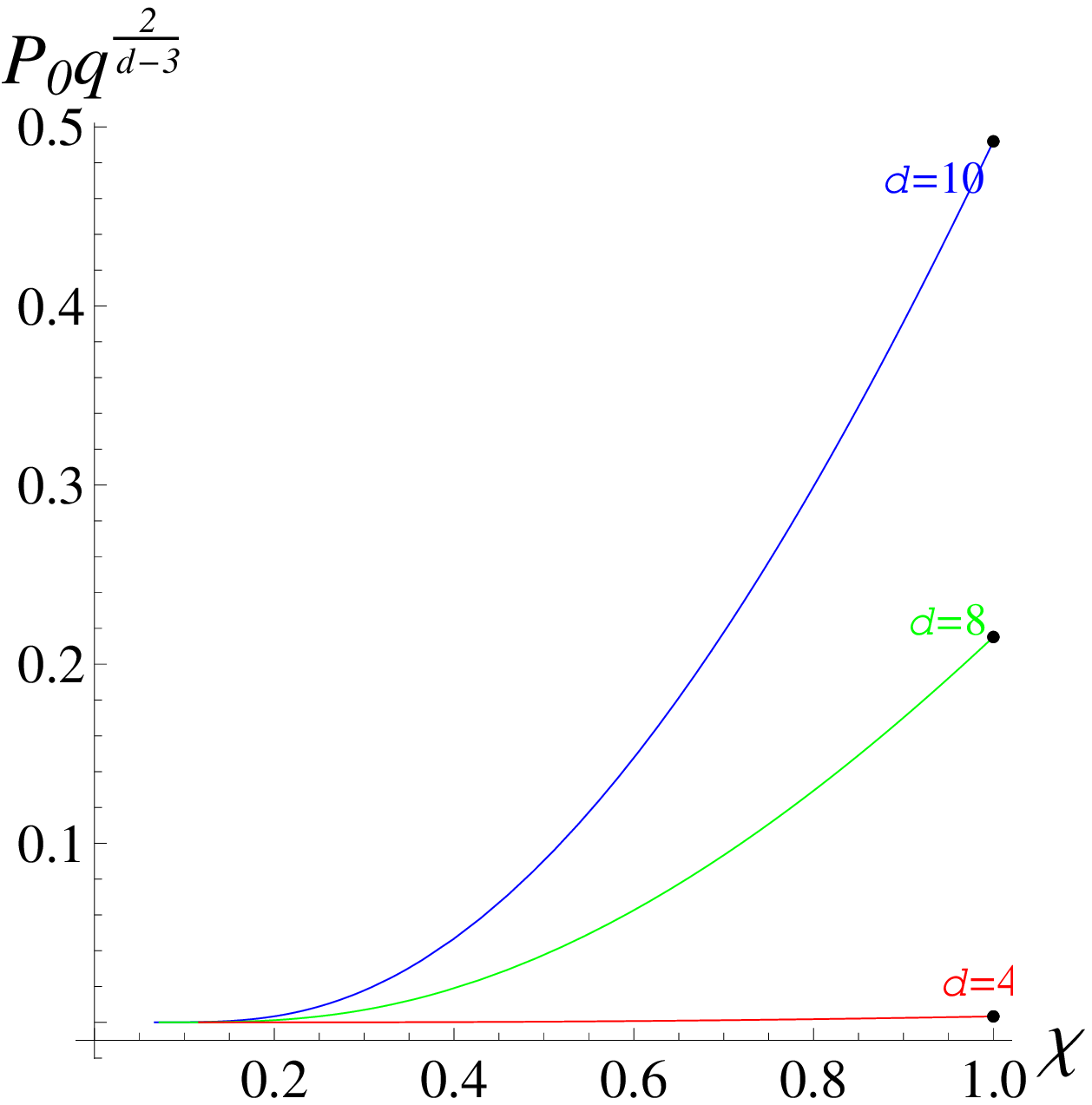}\\
  \caption{\it $P_0 q^{2/(d-3)}-\chi $ phase diagram for $d$-dimensional RN-AdS
black hole with $d=4$,$8$,$10$}\label{P0qchi}
\end{figure}

From (\ref{eq15})-(\ref{eq17}), one can get
\be
\label{eq25}
P_0 =y_3 (x),
\quad
T_0 =y_4 (x),
\ee
in which
\[
y_3 (x)=\frac{d-3}{\pi }\left[ {\frac{1}{\left( {d-2}
\right)x}-\frac{1-x^{2d-5}}{4k^{2d-5}x^{2d-5}\left( {1-x} \right)}\frac{y_2
(x)}{y_1 (x)}} \right]\left( {\frac{y_2 \left( x \right)}{y_1 \left( x
\right)q^2}} \right)^{\frac{1}{d-3}},
\]
\be
\label{eq26}
y_4 (x)=\frac{\left( {d-3} \right)}{\pi }\left[ {\frac{1+x}{\left( {d-2}
\right)x}-\frac{1-x^{2d-4}}{4k^{2d-5}x^{2d-5}\left( {1-x} \right)}\frac{y_2
(x,)}{y_1 (x)}} \right]\left( {\frac{y_2 (x,)}{y_1 (x)q^2}}
\right)^{\frac{1}{2d-6}}.
\ee
From (\ref{eq25})
\be
\label{eq27}
\frac{dP_0 }{dT_0 }=\frac{y_3 '(x)}{y_4 '(x)}
\ee
with $y'(x)=dy/dx$. (\ref{eq27}) stands for the Clapeyron equation of the
thermodynamic system of the $d$-dimensional RN-AdS black holes.

\section{Rotating Black Holes}

\subsection{Thermodynamics}

The AdS rotating black hole solution is given by the Kerr-AdS
metric\cite{Hawking3,Gibbons,Alta1},
\[
ds^2=-\frac{\Delta }{\rho ^2}\left[ {dt-\frac{a\sin ^2\theta }{\Xi }d\phi }
\right]^2+\frac{\rho ^2}{\Delta }dr^2+\frac{\rho ^2}{\Sigma }d\theta ^2
\]
\be
\label{eq28}
+\frac{\Sigma \sin ^2\theta }{\rho ^2}\left[ {adt-\frac{r^2+a^2}{\Xi }d\phi
} \right]^2+r^2\cos ^2\theta d\Omega _{d-4}^2 ,
\ee
where
\[
\Delta =(r^2+a^2)\left( {1+\frac{r^2}{l^2}} \right)-2mr^{5-d},
\quad
\Sigma =1-\frac{a^2}{l^2}\cos ^2\theta ,
\]
\be
\label{eq29}
\Xi =1-\frac{a^2}{l^2},
\quad
\rho ^2=r^2+a^2\cos ^2\theta ,
\ee
and $d\Omega _d^2 $ denotes the metric element on a $d$-dimensional sphere.
The associated thermodynamic quantities are
\[
M=\frac{\omega _{d-2} }{4\pi }\frac{m}{\Xi ^2}\left( {1+\frac{(d-4)\Xi }{2}}
\right),
\quad
J=\frac{\omega _{d-2} }{4\pi }\frac{ma}{\Xi ^2},
\quad
\Omega =\frac{a^2}{l^2}\frac{r_+^2 +l^2}{r_+^2 +a^2},
\]
\[
S=\frac{\omega _{d-2} }{4}\frac{(a^2+r_+^2 )r_+^{d-4} }{\Xi }=\frac{A}{4},
\quad
V=\frac{r_+ A}{d-1}\left[ {1+\frac{a^2}{\Xi }\frac{1+r_+^2 /l^2}{(d-2)r_+^2
}} \right],
\]
\be
\label{eq30}
T=\frac{1}{2\pi }\left[ {r_+ \left( {1+\frac{r_+^2 }{l^2}} \right)\left(
{\frac{1}{a^2+r_+^2 }+\frac{d-3}{2r_+^2 }} \right)-\frac{1}{r_+ }} \right],
\ee
in which $r_+ $ meets $\Delta (r_+ )=0$, that is $\Delta =(r_+ ^2+a^2)\left(
{1+\frac{r_+ ^2}{l^2}} \right)-2mr_+ ^{5-d}=0$. Then
\be
\label{eq31}
\frac{4\pi M\Xi ^2}{\omega _{d-2} \left( {1+\textstyle{{d-4} \over 2}\Xi }
\right)}=m=\frac{1}{2}r_+^{d-5} (r_+^2 +a^2)\left( {1+\frac{r_+^2 }{l^2}}
\right),
\ee
for $r_+ $ in terms of $M$ and $l$. While expressing $a=\varepsilon
l$, we expanded\cite{Alta1}
\be
\label{eq32}
r_+ =\sum\limits_{I=0}^k {r_I \varepsilon ^I}
\ee
to some given order $k$ and solve Eq. (\ref{eq31}) order by order. The first term
yields the relation
\be
\label{eq33}
M=\frac{\omega _{d-2} (d-2)r_0^{d-3} }{16\pi }\left( {1+\frac{r_0^2 }{l^2}}
\right).
\ee
Using Eqs.(\ref{eq6}) and (\ref{eq30}), the equation of $P=P(v,T,J)$ can be obtained
\be
\label{eq34}
P=\frac{T}{v}-\frac{(d-3)}{\pi (d-2)v^2}+\frac{\pi (d-1)16^dJ^2}{4\omega
_{d-2}^2 (d-2)^{2(d-1)}v^{2(d-1)}}+o(\varepsilon ^4),
\ee
where
\be
\label{eq35}
v=\frac{r_0 }{k}+\sum\limits_{I=1} {v_I \varepsilon ^I} .
\ee
While $J=1$ the isotherms are plotted in $P-v$ diagrams at $d=4,\;8,\;10$
respectively, which are shown in Fig.5. The situations of negative pressure
and the thermodynamics unstable region with ${\partial P} \mathord{\left/
{\vphantom {{\partial P} {\partial v>0}}} \right. \kern-\nulldelimiterspace}
{\partial v>0}$ are also existent in the state equation the of Kerr-AdS
black hole. When $T<T_c $, where $T_c $ is the critical temperature of
the Kerr-AdS black hole, unstable thermodynamic state appears, and
while $T<\tilde {T}$, negative pressure emerges. The expression of $\tilde
{T}$ can be derived from (\ref{eq34}) as in (\ref{eq12}), and the corresponding $\tilde
{v}$ is obtained by the way.

\begin{figure}[!htbp]
\center{\subfigure[~$d=4$; $J=1$] {
\includegraphics[angle=0,width=5cm,keepaspectratio]{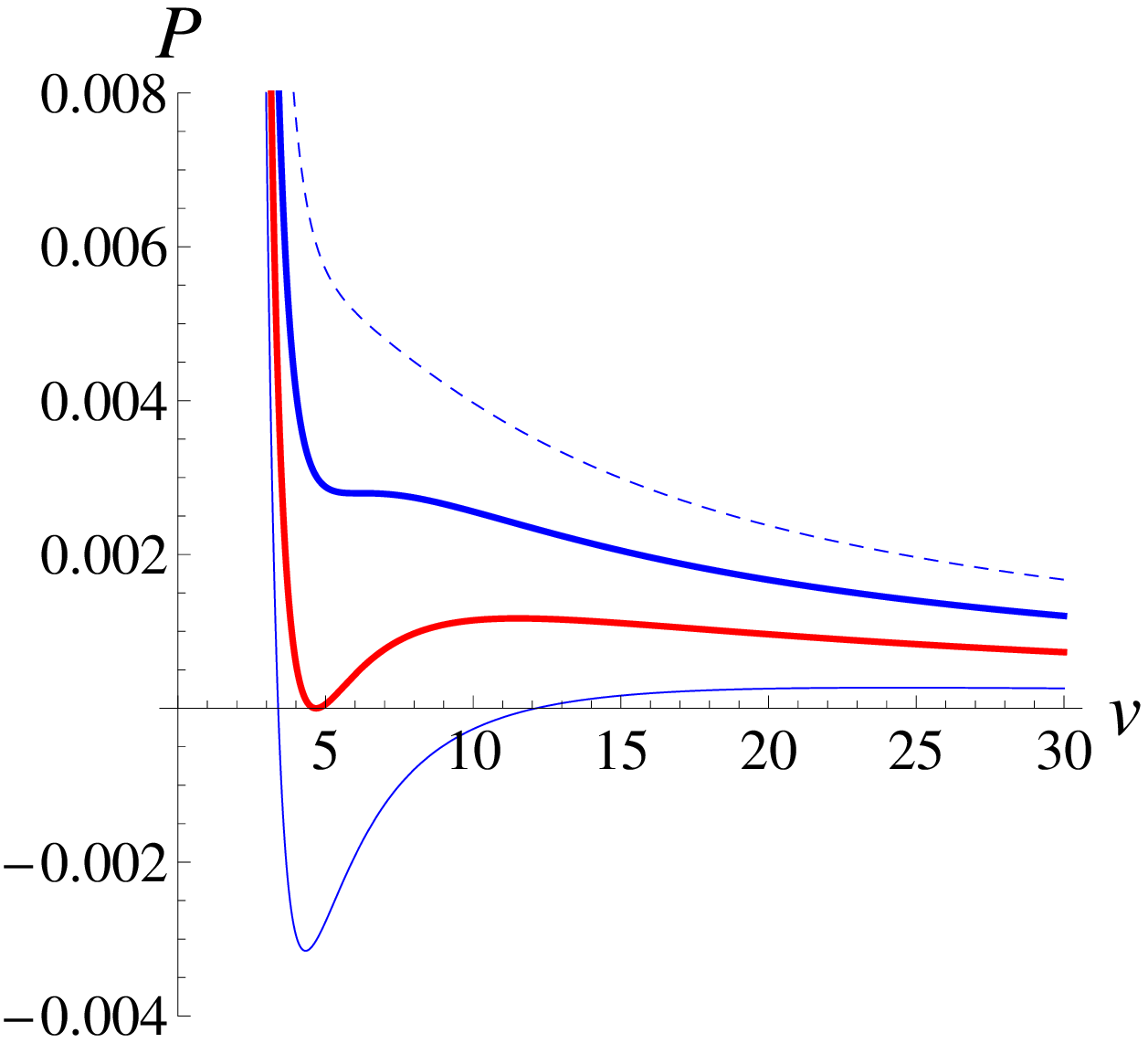}}
\subfigure[~$d=8$; $J=1$] {
\includegraphics[angle=0,width=5cm,keepaspectratio]{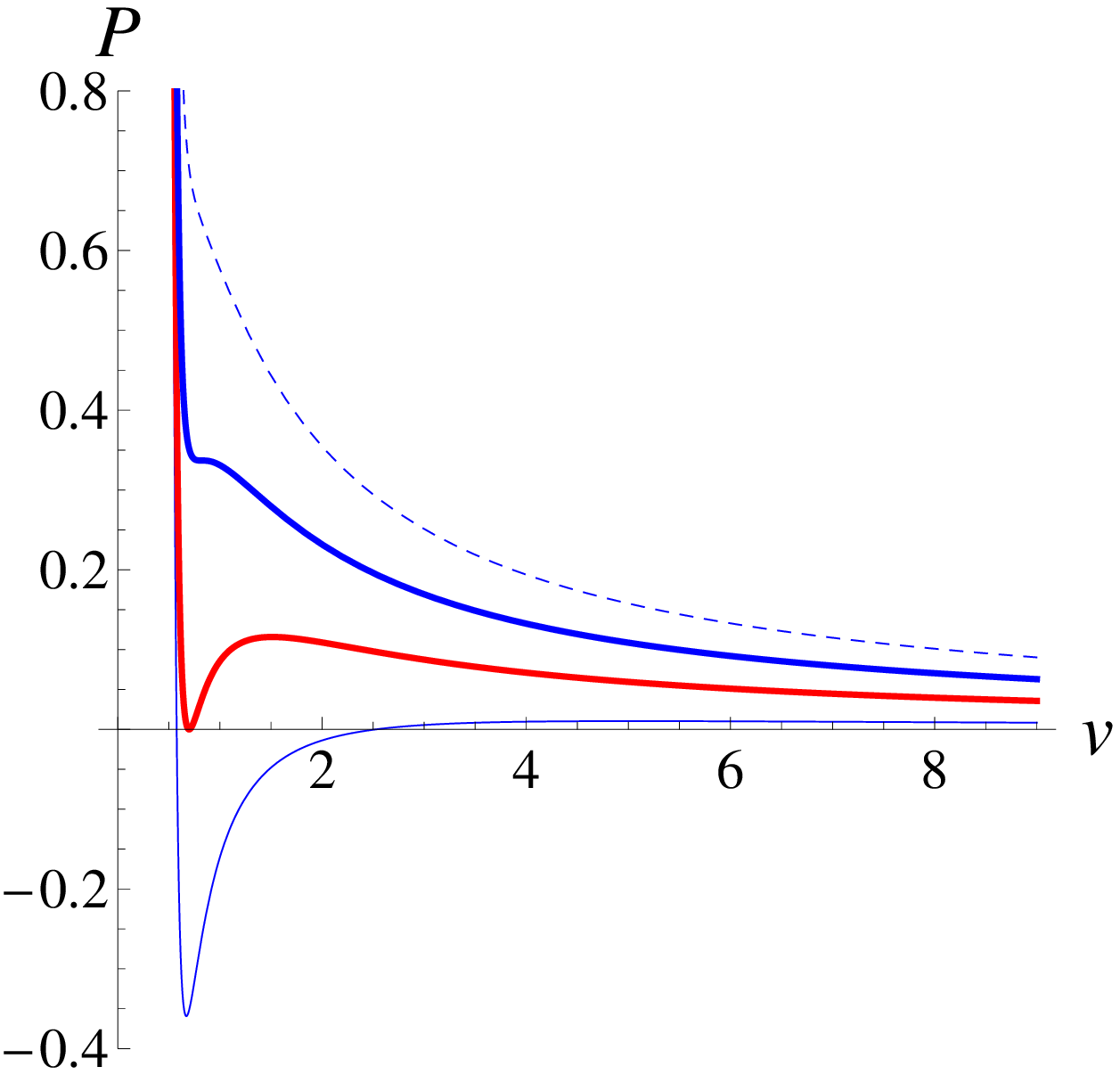}}
\subfigure[~$d=10$; $J=1$] {
\includegraphics[angle=0,width=5cm,keepaspectratio]{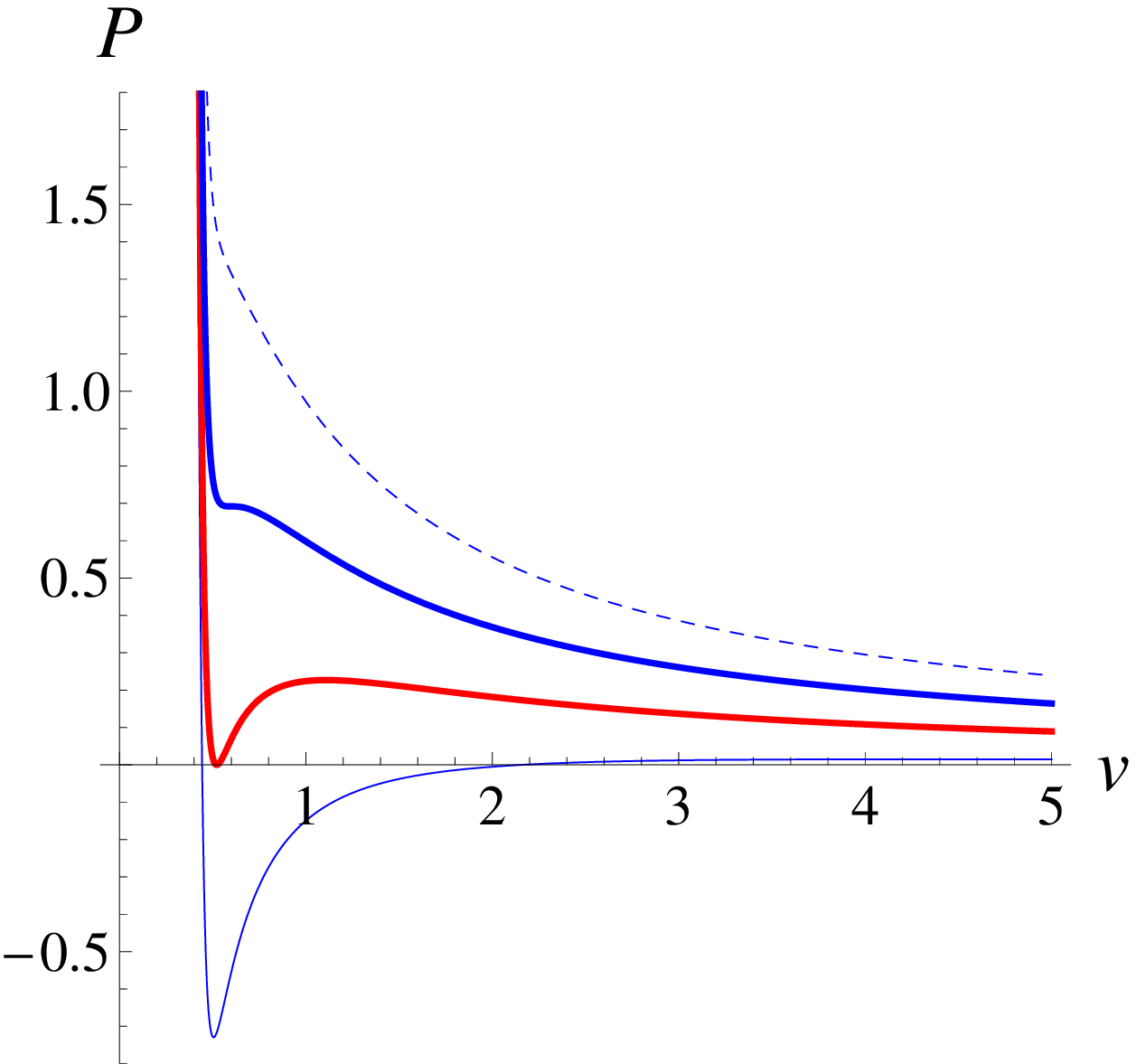}}
\caption[]{\it the isotherms in $P-v$ diagrams . The four curves in
each diagram correspond to $T>T_c $, $T=T_c $, $T=\tilde {T}$,
$T<\tilde {T}$ respectively.}} \label{Pv2}
\end{figure}

\be
\label{eq36}
\tilde{T}=\frac{2(d-3)}{\pi (2d-3)v_0 },
\quad
\tilde {v}^{2d-4}=\frac{\pi ^2(2d-3)(d-1)16^dJ^2}{4\omega _{d-2}^2
(d-3)(d-2)^{2d-3}}.
\ee

\subsection{Maxwell equal area law}

We use Maxwell equal area law to simulate a possible phase transition
process of the $d$-dimensional Kerr-AdS black hole as a thermodynamic system.
We also take $v_1$ and $v_2$ as the specific volumes of the two associated phases in the
phase transition respectively, and $P_0 $ as the constant pressure throughout the
simulated process of the phase transition at constant temperature $T_0 $($T_0 \le T_c $), which can be derived by Maxwell equal area law.

\be
\label{eq37}
P_0 (v_2 -v_1 )=\int\limits_{v_1 }^{v_2 } {Pdv} ,
\ee
which can be rewritten as
\be
\label{eq38}
P_0 (v_2 -v_1 )
=T_0 \ln \left( {\frac{v_2 }{v_1 }} \right)+\frac{(d-3)}{\pi (d-2)}\left(
{\frac{1}{v_2 }-\frac{1}{v_1 }} \right)-\frac{\pi
(d-1)16^dJ^2}{4(2d-3)\omega _{d-2}^2 (d-2)^{2(d-1)}}\left(
{\frac{1}{v_2^{2d-3} }-\frac{1}{v_1^{2d-3} }} \right).
\ee
Combining with
\[
P_0 =\frac{T_0 }{v_1 }-\frac{(d-3)}{\pi (d-2)v_1^2 }+\frac{\pi
(d-1)16^dJ^2}{4\omega _{d-2}^2 (d-2)^{2(d-1)}v_1^{2(d-1)} },
\]
\be
\label{eq39} P_0 =\frac{T_0 }{v_2 }-\frac{(d-3)}{\pi (d-2)v_2^2
}+\frac{\pi (d-1)16^dJ^2}{4\omega_{d-2}^2
(d-2)^{2(d-1)}v_2^{2(d-1)}},
\ee
and setting $x=\frac{v_1 }{v_2 }$, one can get
\[
T_0 =\frac{(d-3)}{\pi (d-2)}\frac{(1+x)}{v_2 x}-\frac{\pi
(d-1)16^dJ^2}{4\omega _{d-2}^2 (d-2)^{2(d-1)}}\frac{1-x^{2\left( {d-1}
\right)}}{v_2^{2d-3} x^{2d-3}\left( {1-x} \right)},
\]
\[
P_0 =\frac{(d-3)}{\pi (d-2)}\frac{1}{v_2^2 x}-\frac{\pi
(d-1)16^dJ^2}{4\omega _{d-2}^2 (d-2)^{2(d-1)}}\frac{1-x^{2d-3}}{v_2^{2d-2}
x^{2d-3}\left( {1-x} \right)},
\]
\be
\label{eq40}
v_2^{2d-4} =J^2\frac{y_5 (x,)}{y_6 (x)},
\ee
where
\[
y_5 (x)=\frac{\pi (d-1)16^d}{2\omega _{d-2}^2 (d-2)^{2(d-1)}}\left(
{(1-x^{2d-2})\ln x+\left[ {\frac{(1-x^{2d-3})}{2d-3}+1-x^{2d-3}}
\right](1-x)} \right),
\]
\be
\label{eq41}
y_6 (x)=\frac{2(d-3)}{\pi (d-2)}x^{2d-4}(1-x)\left( {(1+x)\ln x+2(1-x)}
\right).
\ee
while $x\to 1$, critical thermodynamics quantities can be got (\ref{eq40}),
\[
v_c =\frac{4}{d-2}\left[ {\frac{2^6\pi ^2(2d-3)(d-1)^2J^2}{(d-2)(d-3)\omega
_{d-2}^2 }} \right]^{1/(2d-4)},
\]
\be
\label{eq42}
T_c =\frac{4(d-3)}{\pi (2d-3)v_c },
\quad
P_c =\frac{d-3}{\pi (d-1)v_c^2 }.
\ee
The results are consistent with that in Ref.\cite{Alta1}. As $T_0 =\chi
T_c $ and $\chi \le 1$, we can derive from (\ref{eq40}) and
(\ref{eq42}),
\[
\chi \frac{(d-2)x^{2d-3}}{\pi (2d-3)}\left( {\frac{(d-2)(d-3)\omega _{d-2}^2
}{2^6\pi ^2(2d-3)(d-1)^2}} \right)^{1/(2d-4)}\left( {\frac{y_5 (x)}{y_6
(x)}} \right)^{(2d-3)/(2d-4)}
\]
\be
\label{eq43}
=\frac{(1+x)}{\pi (d-2)}x^{2d-4}\left( {\frac{y_5 (x)}{y_6 (x)}}
\right)-\frac{\pi (d-1)16^d}{4\omega _{d-2}^2 (d-2)^{2(d-1)}\left( {d-3}
\right)}\frac{(1-x^{2(d-1)})}{1-x}.
\ee
So we can solve $x$ at certain $\chi (T_0 )$, then from (\ref{eq40}) we can
obtain the $v_2 $, $P_0 $ and $v_1 =xv_2 $ . Take $J=1$,$d=4;8;10$
respectively. In Fig.6, we plot the isotherms (blue curves) for different
$\chi $ in $P-v$ diagrams and the isobars (red solid lines) which represent
the simulated phase transition processes derived from Maxwell equal area
law. Like that in RN-AdS black hole the processes become
shorter with increasing temperatures until it turns into a point at a certain
temperature, which is the critical temperature, and the point is the
critical point of the black hole thermodynamic system. Every connected solid
curve represents a processes of state change and phase transition,
where no negative pressure and thermodynamic unstable state appear. The
boundary of the two-phase coexistence region is delineated by
the dot-dashed line.

\begin{figure}
  \centering
  \includegraphics[width=3in]{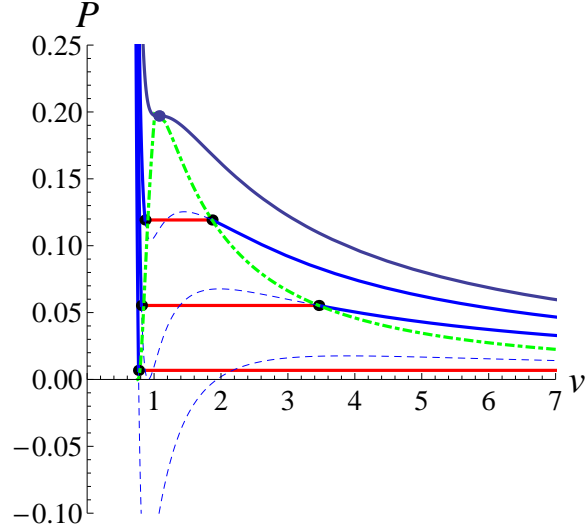}\\
  \caption{\it The simulated phase transition(red solid lines) and the boundary of
two phase coexistence (green dot-dashed curve) on the base of
isotherms in $P-v$ diagram for Kerr-AdS black hole with $d=8$,
$J=5$.}\label{coexist2}
\end{figure}

In Table 2, we calculate $x$, $v_2 $, $P_0 $ as $\chi
=0.1,\;0.3,\;0.5,\;0.7,\;0.9$, $J=0.2,\;0.5,\;1$, and $d=4,\;8,\;10$
respectively to find the influence of these parameters on the simulated
phase transition process and the two-phase coexistence region.

\begin{table}[htbp]
\caption{\it The parameters of two phase coexistence for Kerr-AdS black hole}
\begin{center}
\begin{tabular}{|p{20pt}|p{20pt}|p{40pt}|p{40pt}|p{40pt}|p{40pt}|p{40pt}|p{40pt}|p{40pt}|p{40pt}|p{40pt}|}

\hline & & \multicolumn{3}{|p{120pt}|}{$d=4$} &
\multicolumn{3}{|p{120pt}|}{$d=8$} &
\multicolumn{3}{|p{120pt}|}{$d=10$}  \\

\hline $J$& $\chi $& $x$& $v_2 $& $P_0 $& $x$& $v_2 $& $P_0 $& $x$&
$v_2 $& $P_0 $ \\

\hline \raisebox{-6.00ex}[0cm][0cm]{$0.2$}
 & 0.1& 1.98E{-5}& 7.21E{4}& 1.28E{-7}& 2.56E{-4}& 1.71E{3}& 4.56E{-5}& 3.95E{-4}& 913.& 1.17E{-4} \\

\cline{2-11}
 & 0.3& 0.0166& 90.9& 2.86E{-4}&0.0428& 10.5& 0.0199& 0.0506& 7.29& 0.0389 \\

\cline{2-11}
 & 0.5& 0.0859& 18.8& 0.00201& 0.162& 2.87& 0.104& 0.182& 2.09& 0.193 \\

\cline{2-11}
 & 0.7& 0.218& 8.12& 0.00556& 0.345& 1.42& 0.253& 0.375& 1.05& 0.460 \\

\cline{2-11}
 & 0.9& 0.475& 4.36& 0.0108& 0.617& 0.862& 0.457& 0.650& 0.654& 0.824 \\

\hline \raisebox{-6.00ex}[0cm][0cm]{$0.5$}
 & 0.1& 1.98E{-5}& 1.14E{5}& 5.12E{-8}& 2.56E{-4}& 1.99E{3}& 3.36E{-5}& 3.95E{-4}& 1.02E{3}& 9.31E{-5} \\

\cline{2-11}
 & 0.3& 0.0166& 144.& 1.14E{-4}& 0.0428& 12.2& 0.0146& 0.0506& 8.17& 0.0309 \\

\cline{2-11}
 & 0.5& 0.0859& 29.8& 8.03E{-4}& 0.162& 3.34& 0.0763& 0.182& 2.34& 0.153 \\

\cline{2-11}
 & 0.7& 0.218& 12.8& 0.00222& 0.345& 1.65& 0.186& 0.375& 1.18& 0.367 \\

\cline{2-11}
 & 0.9& 0.475& 6.89& 0.00432& 0.617& 1.00& 0.337& 0.650& 0.732& 0.655 \\

\hline \raisebox{-6.00ex}[0cm][0cm]{$1$}
 & 0.1& 1.98E{-5}& 1.61E{5}& 2.56E{-8}& 2.56E{-4}& 2.23E{3}& 2.67E{-5}& 3.95E{-4}& 1.12E{3}& 7.83E{-5} \\

\cline{2-11}
 & 0.3& 0.0166& 203.& 5.72E{-5}& 0.0428& 13.7& 0.0116& 0.0506& 8.91& 0.0260 \\

\cline{2-11}
 & 0.5& 0.0859& 42.1& 4.01E{-4}& 0.162& 3.75& 0.0606& 0.182& 2.55& 0.129 \\

\cline{2-11}
 & 0.7& 0.218& 18.2& 0.00111& 0.345& 1.86& 0.148& 0.375& 1.29& 0.308 \\

\cline{2-11}
 & 0.9& 0.475& 9.75& 0.00216& 0.617& 1.13& 0.267& 0.650& 0.799& 0.551 \\
\hline
\end{tabular}
\label{tab2}
\end{center}
\end{table}

From Table 2, one can see that $x$ is unrelated to $J$, but incremental with
the increase of $\chi $/$d$ at certain $d$/$\chi$ . $v_2 $
decreases with increasing $\chi$/ $d$ and increases with the incremental $J$
when the other parameters are fixed respectively. Similarly $P_0 $ increases with the incremental $\chi$/$d$
and decreases with increasing $J$ while the other parameters are given
respectively.

\subsection{the relation of phase transition pressure to temperature}

From (\ref{eq39}) and $T_0=\chi T_c $, we get
\[
P_0 J^{2/(d-2)}\left( {\frac{y_5 (x)}{y_6 (x)}} \right)^{(d-1)/(d-2)}=\chi
\frac{(d-3)(d-2)}{\pi (2d-3)}\left( {\frac{y_5 (x)}{y_6 (x)}}
\right)^{(2d-3)/(2d-4)}
\]
\be
\label{eq44}
\left[ {\frac{(d-2)(d-3)\omega _{d-2}^2 }{2^6\pi
^2(2d-3)(d-1)^2}} \right]^{1/(2d-4)}
-\frac{(d-3)}{\pi (d-2)}\left( {\frac{y_5 (x)}{y_6 (x)}} \right)+\frac{\pi
(d-1)16^d}{4\omega _{d-2}^2 (d-2)^{2(d-1)}}.
\ee
Combining (\ref{eq43}), (\ref{eq44}) and $T_0=\chi T_c$, we plot the $P_0
J^{2/(d-2)}-\chi $ diagram and $P-T$ diagrams in Fig.7 and Fig.8.

\begin{figure}
  \centering
  \includegraphics[width=3in]{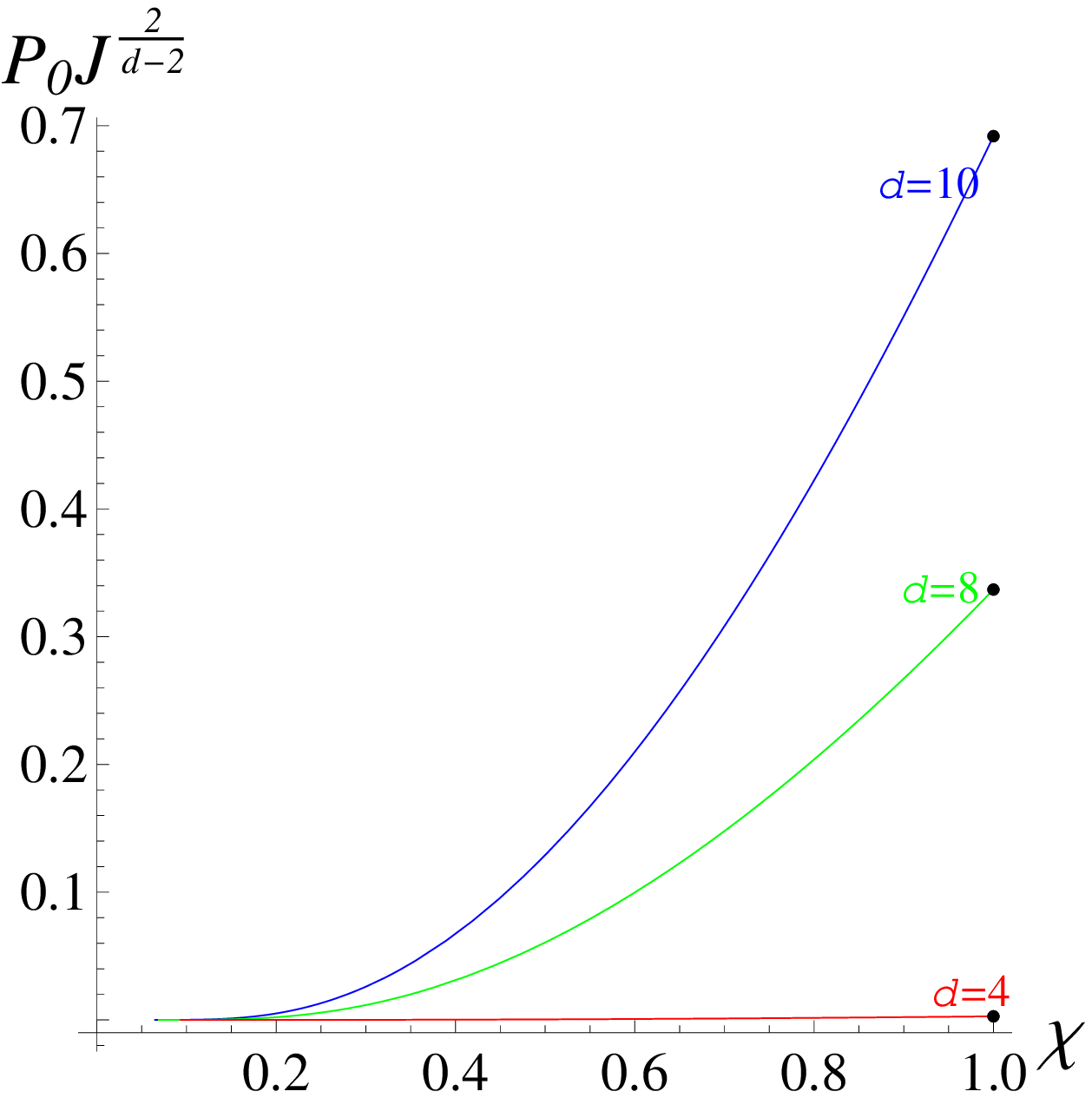}\\
  \caption{\it $P_0 J^{2/(d-2)}-\chi $ phase diagram for $d$-dimensional Kerr-AdS
black hole with $d=4$,$8$,$10$.}\label{P0Jchi}
\end{figure}

\begin{figure}[!htbp]
\center{\subfigure[~$d=4$; $J=1$] {
\includegraphics[angle=0,width=5cm,keepaspectratio]{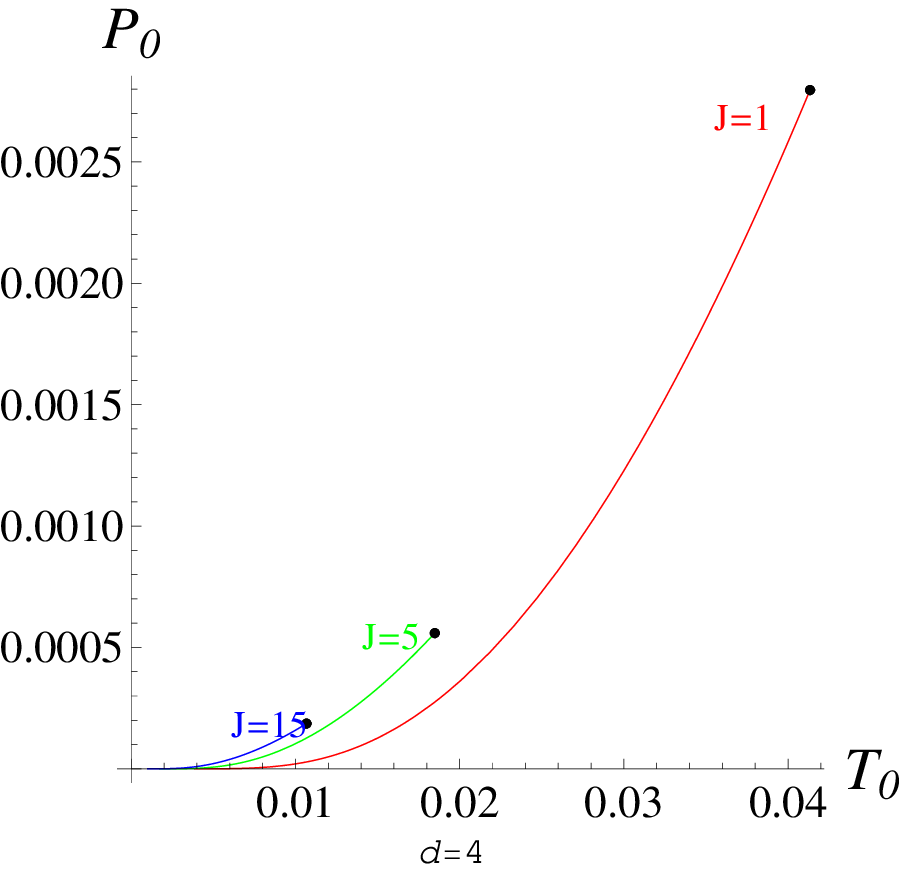}}
\subfigure[~$d=8$; $J=1$] {
\includegraphics[angle=0,width=5cm,keepaspectratio]{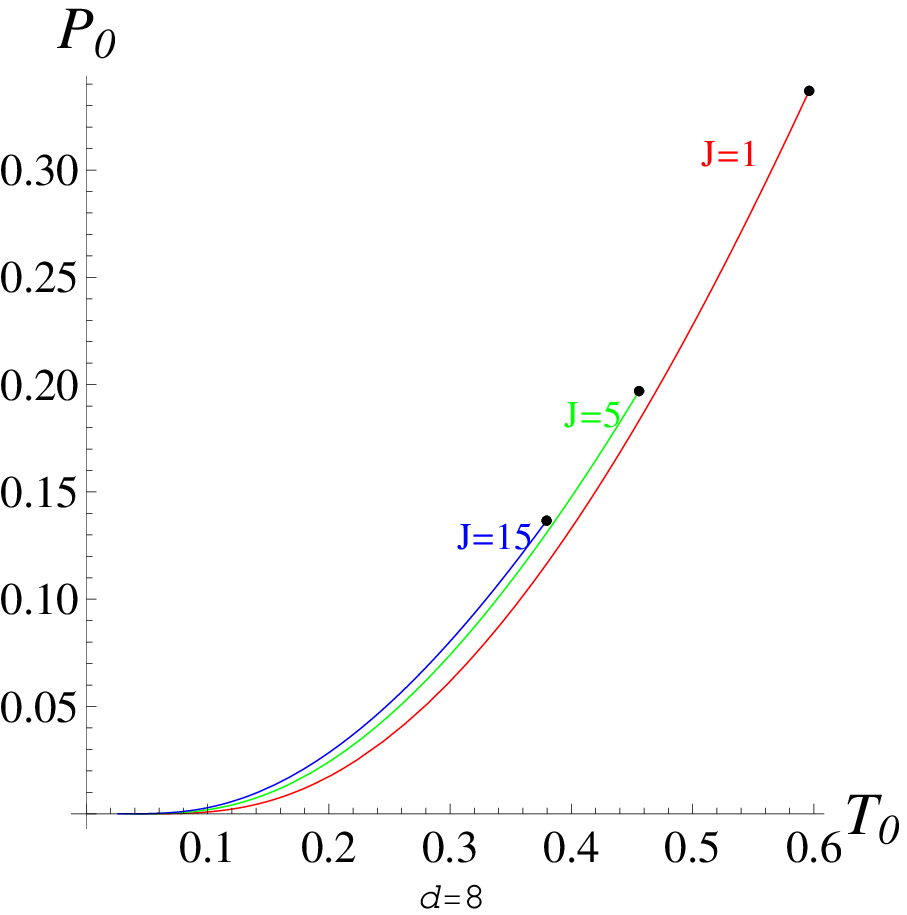}}
\subfigure[~$d=10$; $J=1$] {
\includegraphics[angle=0,width=5cm,keepaspectratio]{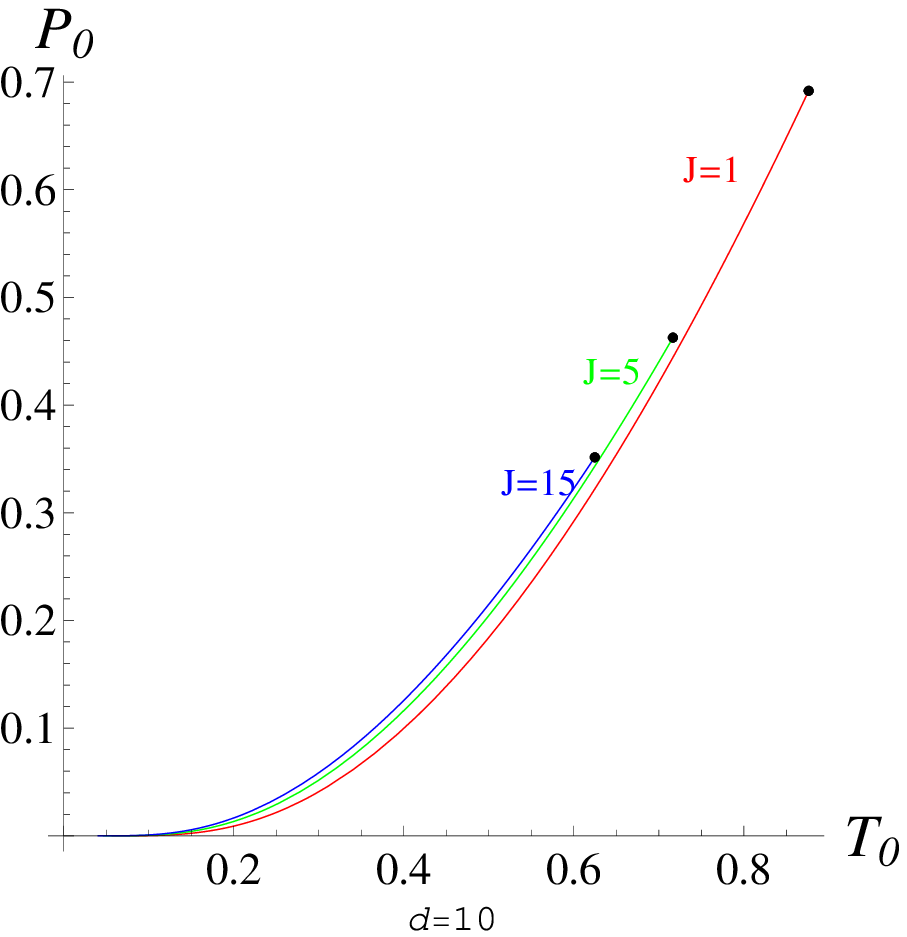}}
\caption[]{\it $P-T$ phase diagrams at fixed $J$ and $d$ for
$d$-dimensional Kerr-AdS black hole.}} \label{PT2}
\end{figure}

From (\ref{eq40}), we get
\be
\label{eq45}
P_0 =y_7 (x),
\quad
T_0 =y_8 (x),
\ee
where
\[
y_7 \left( x \right)=J^{\frac{-2}{d-2}}\left( {\frac{y_6 (x)}{y_5 (x)}}
\right)^{\frac{d-1}{d-2}}\left( {\frac{(d-3)}{\pi (d-2)x}\frac{y_5 (x)}{y_6
(x)}-\frac{\pi (d-1)16^d}{4\omega _{d-2}^2
(d-2)^{2(d-1)}}\frac{1-x^{2d-3}}{x^{2d-3}\left( {1-x} \right)}} \right),
\]
\[
y_8 (x)=J^{-\frac{1}{d-2}}\left( {\frac{y_6 (x)}{y_5 (x)}}
\right)^{\frac{2d-3}{2d-4}}\left( {\frac{(d-3)}{\pi
(d-2)}\frac{(1+x)}{x}\frac{y_5 (x)}{y_6 (x)}-\frac{\pi (d-1)16^d}{4\omega
_{d-2}^2 (d-2)^{2(d-1)}}\frac{1-x^{2\left( {d-1} \right)}}{x^{2d-3}\left(
{1-x} \right)}} \right).
\]
Then
\be
\label{eq46}
\frac{dP_0 }{dT_0 }=\frac{y_7 '(x)}{y_8 '(x)},
\ee
which is the Clapeyron equation of the Kerr-AdS black hole.

We can see from Fig.7, $P_0 \ge 0$, and $P_0 \to 0$ as $\chi \to 0(T\to
0)$. So there is no the situation of negative pressure and the thermodynamic
unstable states in the simulated phase transition processes of
$d$-dimensional Kerr-AdS black hole, as that of the $d$-dimensional RN-AdS
black hole.

\section{Discussions}

In this paper we take AdS black holes as thermodynamic systems and find the
state equations in some region yield negative pressure
and thermodynamic unstable states with ${\partial P} \mathord{\left/
{\vphantom {{\partial P} {\partial v}}} \right. \kern-\nulldelimiterspace}
{\partial v}>0$. Comparing with phase transition in a usual liquid-gas system
and using Maxwell equal area law, we simulate the phase transition process
in AdS black holes. From Fig.2 and Fig.6,
one can see that, for both the $d$-dimensional RN-AdS black hole and
the $d$-dimensional Kerr-AdS black hole, as $T<T_c $ a part of each isotherm could be
replaced with an isobar, which means phase transition occurs and two phases
coexist in the section. The simulated phase transition belongs to the first
order phase transition, and as $T\to T_c $ the phase transition process
shortens gradually until it become a point at $T=T_c $, where the phase
transition is the second order phase transition. The dimension $d$ and the
electric charge $q$ (the angular momentum $J$)
also have effect on the simulated phase transition process. The larger the dimension
$d$, the greater the change of specific volume before and after the phase
transition. The larger the parameter $q(J)$, the lower the pressure $P_0 $,
in the simulated phase transition.

Due to lack of the knowledge of chemical potential, the $P-T$ curve for two phase
coexistence state of an usual thermodynamic system are obtained by
experiment. However the slope of the curves can be calculated from the
Clapeyron equation,
\be
\label{eq47}
\frac{dP}{dT}=\frac{L}{T(v^\beta -v^\alpha )},
\ee
in which $L=T(s^\beta -s^\alpha )$, $v^\alpha $ and $v^\beta $are the molar
volumes of phase $\alpha $ and phase $\beta $ respectively, and the
Clapeyron equation accords with experiment result, which directly verifies its
thermodynamic correctness.

In a general thermodynamic system, the phase transition means the change of
organization structure of substance (including the changes of position and
the orientation of atom, ion, and electron) and a different phase appear.
Phase transition is a physical process, and it does not involve the chemical
reaction, so the chemical compositions do not change in the process.

Appropriate theoretical interpretation to the phase structure of AdS black
hole thermodynamic system can help to know more about black hole
thermodynamic properties, such as entropy, temperature and heat capacity, and
it is significant for improving a self-consistent black hole thermodynamic
theory. This paper displays the $P-T$ diagrams for the two phase coexistence
state and the Clapeyron equations of the black holes, which provide theoretical basis for investigation of phase
transition and phase structure of black holes and help to explore the theory of gravity.

\begin{acknowledgments}\vskip -4mm
This work is supported by NSFC under Grant
Nos.(11475108,11175109;11075098), by the Shanxi Datong University
doctoral Sustentation Fund Nos. 2011-B-03, China, Program for the
Innovative Talents of Higher Learning Institutions of Shanxi, and
the Natural Science Foundation for Young Scientists of Shanxi
Province,China (Grant No.2012021003-4).

\end{acknowledgments}

\end{document}